# THE QUANTUM HANDSHAKE EXPLORED

**JOHN G CRAMER**, Department of Physics, University of Washington
email: jcramer@uw.edu

We discuss the transactional interpretation of quantum mechanics, apply it to several counter-intuitive quantum optics experiments (two-slit, quantum eraser, trapped atom, ...), and describe a mathematical model that shows in detail how transactions form.

## 1  QUANTUM ENTANGLEMENT AND NONLOCALITY

Quantum mechanics, our standard theoretical model of the physical world at the smallest scales of energy and size, differs from the classical mechanics of Newton that preceded it in one very important way. Newtonian systems are always local. If a Newtonian system breaks up, each of its parts has a definite and well-defined energy, momentum, and angular momentum, parceled out at breakup by the system while respecting conservation laws. After the component subsystems are separated, the properties of any subsystem are completely independent and do not depend on those of the other subsystems.

On the other hand, quantum mechanics is nonlocal, meaning that the component parts of a quantum system may continue to influence each other, even when they are well separated in space and out of speed-of-light contact. This characteristic of standard quantum theory was first pointed out by Albert Einstein and his colleagues Boris Podolsky and Nathan Rosen (EPR) in 1935, in a critical paper[1] in which they held up the discovered nonlocality as a devastating flaw that, it was claimed, demonstrated that the standard quantum formalism must be incomplete or wrong. Einstein called nonlocality "spooky actions at a distance". Schrödinger followed on the discovery of quantum nonlocality by showing in detail how the components of a multi-part quantum system must depend on each other, even when they are well separated [2].

Beginning in 1972 with the pioneering experimental work of Stuart Freedman and John Clauser[3], a series of quantum-optics EPR experiments testing Bell inequality violations [4] and other aspects of entangled quantum systems were performed. This body of experimental results can be taken as a demonstration that, like it or not, both quantum mechanics and the underlying reality it describes are intrinsically nonlocal. Einstein's spooky actions-at-a-distance are really out there in the physical world, whether we understand and accept them or not.

How and why is quantum mechanics nonlocal? Nonlocality comes from two seemingly conflicting aspects of the quantum formalism: (1) energy, momentum, and angular momentum, important properties of light and matter, are conserved in all quantum systems, in the sense that, in the absence of external forces and torques, their net values must remain unchanged as the system evolves, while (2) in the wave functions describing quantum systems, as required by Heisenberg's uncertainty principle[5], the conserved quantities may be indefinite and unspecified and typically can span a large range of possible values. This non-specificity persists until a measurement is made that "collapses" the wave function and fixes the measured quantities with specific values. These seemingly inconsistent requirements of (1) and (2) raise an important question: how can the wave functions describing the separated members of a system of particles, which may be light-years apart, have arbitrary and unspecified values for the conserved quantities and yet respect the conservation laws when the wave functions are collapsed?

This paradox is accommodated in the formalism of quantum mechanics because the quantum wave functions of particles are entangled – the term coined by Schrödinger [2] to mean that even when the wave functions describe system parts that are spatially separated and out of light-speed contact, the separate wave functions continue to depend on each other and cannot be separately specified. In particular, the conserved quantities in the system's parts (even though individually indefinite) must always add up to the values possessed by the overall quantum system before it separated into parts.

How could this entanglement and preservation of conservation laws possibly be arranged by Nature? The mathematics of quantum mechanics gives us no answers to this question: it only insists that the wave functions of separated parts of a quantum system do depend on each other. Theorists prone to abstraction have found it convenient to abandon the three-dimensional universe and describe such quantum systems as residing in a many-dimensional Hilbert hyperspace in which the conserved variables form extra dimensions and in which the interconnections between particle wave functions are represented as allowed sub-regions of the overall hyperspace. That has led to elegant mathematics, but it provides little assistance in visualizing what is really going on in the physical world.

Then, how is this behavior possible? The transactional interpretation of quantum mechanics provides the answer.

# 2 THE TRANSACTIONAL INTERPRETATION OF QUANTUM MECHANICS

The Transactional Interpretation of quantum mechanics [6-11], inspired by the structure of the quantum wave mechanics formalism itself, views each quantum event as a Wheeler-Feynman[12] "handshake" or "transaction" process extending across spacetime that involves the exchange of advanced and retarded quantum wave functions to enforce the conservation of certain quantities (energy, momentum, angular momentum, etc.). It asserts that each quantum transition forms in four stages: (1) emission, (2) response, (3) stochastic choice, and (4) repetition to completion.

The first stage of a quantum event is the emission of an "offer wave" by the "source", which is the object supplying the quantities transferred. The offer wave is the time-dependent retarded quantum wave function $\psi$, as used in standard quantum mechanics. It spreads through spacetime until it encounters the "absorber", the object receiving the conserved quantities.

The second stage of a quantum event is the response to the offer wave by any potential absorber (there may be many in a given event). Such an absorber produces an advanced "confirmation wave" $\psi*$, the complex conjugate of the quantum offer wave function $\psi$. The confirmation wave travels in the reverse time direction and arrives back to the source at precisely the instant of emission with an amplitude given by $\psi\psi*$.

The third stage of a quantum event is the stochastic choice that the source exercises in selecting one of the many received confirmations. The strengths $\psi\psi*$ of the advanced-wave "echoes" determine which transaction forms in a linear probabilistic way.

The final stage of a quantum event is the repetition to completion of this process by the source and selected absorber, each perturbed by the other in an unstable configuration that avalanches to completion, reinforcing the selected transaction with multiple wave exchanges until the conserved quantities are transferred, the states stabilize, and the potential quantum event becomes a real event.

Here we summarize the principal elements of the Transactional Interpretation, structured in order to contrast it with the Copenhagen Interpretation:

- The fundamental quantum mechanical interaction is taken to be the transaction. The state vector $\psi$ of the quantum mechanical formalism is a physical wave with spatial extent and is identical to the initial "offer wave" of the transaction. The complex conjugate of the state vector $\psi*$ is also a physical wave and is identical to the subsequent "confirmation wave" of the transaction. The particle (photon, electron, etc.) and the collapsed state vector are identical to the completed transaction. The transaction may involve a single emitter and absorber and two vertices or multiple emitters and absorbers and many vertices, but is only complete when appropriate quantum boundary conditions are satisfied at all vertices, i.e., loci of emission and absorption. Particles transferred have no separate identity independent from the satisfaction of the boundary conditions at the vertices.

- The correspondence of "knowledge of the system" with the state vector y is a fortuitous but deceptive consequence of the transaction, in that such knowledge must necessarily follow and describe the transaction.

- Heisenberg's Uncertainty Principle [5] is a consequence of the fact that a transaction in going to completion is able to project out and localize only one of a pair of conjugate variables (e.g., position or momentum) from the offer wave, and in the process it delocalizes the other member of the pair, as required by the mathematics of Fourier analysis. Thus, the Uncertainty Principle is a consequence of the transactional model and is not a separate assumption.

- Born's Probability Rule [13] is a consequence of the fact that the magnitude of the "echo" received by the emitter, which initiates a transaction in a linear probabilistic way, has strength $\mathbf{P} = \psi\psi*$. Thus, Born's Probability Rule is a consequence of the transactional model and is not a separate assumption of the interpretation.

- All physical processes have equal status, with the observer, intelligent or otherwise, given no special status. Measurement and measuring apparatus have no special status, except that they happen to be processes that connect and provide information to observers.

- Bohr's "wholeness" of measurement and measured system exists, but is not related to any special character of measurements but rather to the connection between emitter and absorber through the transaction.

- Bohr's "complementarity" between conjugate variables exists, but like the Uncertainty Principle is just a manifestation of the requirement that a given transaction going to completion can project out only one of a pair of conjugate variables, as required by the mathematics of Fourier analysis.

- Resort to the positivism of "don't-ask-don't-tell" is unnecessary and undesirable. A distinction is made between observable and inferred quantities. The former are firm predictions of the overall theory and may be subjected to experimental verification. The latter, particularly those that are complex quantities, are not verifiable and are useful only for visualization, interpretational, and pedagogical purposes. It is assumed that both kinds of quantities must obey conservation laws, macroscopic causality conditions, relativistic invariance, etc.

In summary, the Transactional Interpretation explains the origin of the major elements of the Copenhagen Interpretation while avoiding their paradoxical implications. It drops the positivism of the Copenhagen Interpretation as unnecessary, because the positivist curtain is no longer needed to hide the nonlocal backstage machinery.

It should also be pointed out that giving some level of objective reality to the state vector colors all of the other elements of the interpretation. Although in the Transactional Interpretation, the Uncertainty Principle and the statistical interpretation are formally the same as in the Copenhagen Interpretation, their philosophical implications, about which so much has been written from the Copenhagen viewpoint, may be rather different.

J.C. CramerThe Transactional Interpretation offers the possibility of resolving all of the many interpretational paradoxes that quantum mechanics has accumulated over the years. Many of these are analyzed in reference [6], the publication in which the Transactional Interpretation was introduced. Here we will not attempt to deal with all of the paradoxes. We will instead focus on the interpretational problems associated with quantum nonlocality and entanglement.

## 3  APPLYING THE TRANSACTIONAL INTERPRETATION TO QUANTUM PARADOXES

### 3.1  Einstein's Bubble *Gedankenexperiment* (1927)

Quantum nonlocality is one of the principal counterintuitive aspects of quantum mechanics. Einstein's "spooky action-at-a-distance" is a real feature of quantum mechanics, but the quantum formalism and the orthodox Copenhagen Interpretation provide little assistance in understanding nonlocality or in visualizing what is going on in a nonlocal process. The Transactional Interpretation provides the tools for doing this. Perhaps the first example of a nonlocality paradox is the Einstein's bubble paradox was proposed by Albert Einstein at the 5th Solvay Conference in 1927 [14,15].

A source emits a single photon isotropically, so that there is no preferred emission direction. According to the Copenhagen view of the quantum formalism, this should produce a spherical wave function ψ that expands like an inflating bubble centered on the source. At some later time, the photon is detected, and, since the photon does not propagate further, its wave function bubble should "pop", disappearing instantaneously from all locations except the position of the detector. Einstein asked how the parts of the wave function away from the detector could "know" that they should disappear, and how it could be arranged that only a single photon was always detected when only one was emitted?

At the 5th Solvay Conference, Werner Heisenberg [15] dismissed Einstein's bubble paradox by asserting that the wave function cannot be depicted as a real object moving through space, as Einstein had implicitly assumed, but instead is a mathematical representation of the knowledge of some observer who is watching the process. Until detection, the observer knows nothing about the location of the emitted photon, so the wave function must be spherical, distributed over the 4π solid angle to represent his ignorance. However, after detection the location of the photon is known to the observer, so the wave function "collapses" and is localized at the detector. One photon is detected because only one photon was emitted.

The Transactional Interpretation provides an alternative explanation, one that permits the wave function to be, in some sense, a real object moving through space rather than an esoteric representation of knowledge. This is illustrated in Fig. 1. The offer wave ψ from the source indeed spreads out as a spherical wave front and eventually encounters the detector on the right. The detector responds by returning to the source a confirmation wave ψ∗. Other detectors (i.e., potential absorbers) also return confirmation waves, but the source, randomly weighted by the ψψ∗ echoes from the potential absorbers, selects the detector on the right to form a transaction. The transaction forms between source and detector, and one ℏω photon's worth of energy is transferred from the source to the detector. The formation of this particular transaction, satisfying the source boundary condition that only one photon is emitted, prevents the formation of any other transaction to another possible photon absorber, so only one photon is detected. This is an illustration of a simple two-vertex transaction in which the transfer of a single photon is implemented nonlocally. It avoids Heisenberg's assertion that the mathematical solution to a simple second-order differential equation involving momentum, energy, time, and space has somehow become a map of the mind, deductions, and knowledge of a hypothetical observer.

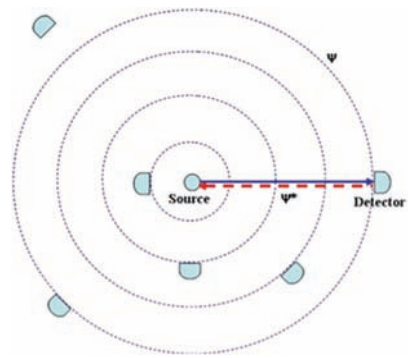

**Fig. 1:** Schematic of the transaction involved in the Einstein's bubble paradox. The offer wave ψ (blue/solid) forms a spherical wave front, reaching the detector on the right and causing it to return a confirmation wave ψ∗ (red/dashed), so that a transaction forms and one photon's worth of energy ℏω is transferred. Other detectors also return confirmation waves, but the source has randomly selected the detector on the right for the transaction.

In this context, we note that there is a significant (but untestable) difference between Heisenberg's knowledge interpretation and the Transactional Interpretation as to whether the outgoing state vector or offer wave changes, collapses, or disappears at the instant when knowledge from a measurement is obtained. The knowledge interpretation would lead us to expect, without any observational evidence and with some conflict with special relativity, that Einstein's bubble "pops" when the detector registers the arrival of a photon and that other parts of the outgoing wave disappear at that instant. The bubble needs to pop in the knowledge interpretation because the state of knowledge changes, and also because this prevent multiple photon detections from a single photon emission.

In the analogous description by the Transactional Interpretation, the parts of the offer wave away from the detection site, because they represent only the *possibility* of a quantum event, do not disappear, but instead continue to propagate to more distant potential detection sites. These sites return confirmation echoes that compete with the echo from the detector of interest for transaction formation. The consequence of this difference is that the TI does not have to explain how wave functions can change in mid-flight, how the absence of a detection can change a propagating wave function, or what "instantaneous disappearence" means in the context of special relativity.

### 3.2  Young's Two-Slit Experiment (1893)

Thomas Young (1773–1829) presented the results of his two-slit experiment to the Royal Society of London on November 24, 1803. A century and a half later, Richard Feynman [16] described Young's experiment as "a phenomenon that is impossible… to explain in any classical way, and that has in it the heart of quantum mechanics. In reality, it contains the only (quantum) mystery."

The experimental arrangement of Young's two-slit experi-

374

ment is shown in Fig. 2. Plane waves of light diffract from a small aperture in screen **A**, pass through two slits in screen **B**, and produce an interference pattern in their overlap region on screen **C**. The interference pattern is caused by the arrival of light waves at screen **C** from the two slits, with a variable relative phase because the relative path lengths of the two waves depends on the location on screen **C**. When the path lengths are equal or differ by an integer number of light wavelengths **λ**, the waves add coherently (constructive interference) to produce an intensity maximum. When the path lengths differ by an odd number of half-wavelengths **λ/2**, the waves subtract coherently to zero (destructive interference) and produce an intensity minimum.

One can "turn off" this interference pattern by making the two paths through slits distinguishable. In this case, the "comb" interference pattern is replaced by a broad diffraction "bump" distribution, as shown by the green/dashed line at **C** in Fig. 2. This might be accomplished by arranging for the waves on the two paths to be in different polarization states, thereby "labeling" the wave paths with polarization. For example, one could use a light source that produces vertically polarized light, and one could place behind one slit a small optical half-wave plate, shown in Fig. 2 behind the upper slit at **B**, set to rotate vertical to horizontal polarization. This would eliminate the previously observed two-slit interference pattern, because the light waves arriving at screen **C** from the two slits are now in distinguishable polarization states, with the waves from the lower slit vertically polarized and waves from the upper slit horizontally polarized. The intensities of the waves will now add instead of their amplitudes, and there can be no destructive cancellation. This interference suppression occurs even if no polarization is actually measured at **C**.

In the 19th century Young's experiment was taken as conclusive proof that light was a wave and that Newton's earlier depiction of light as a particle was incorrect. Einstein's 1905 explanation of the photoelectric effect as caused by the emission of photon particles of light cast doubt on this view. In 1909, a low-intensity double slit experiment performed by Sir Geoffrey Taylor [17] demonstrated that the same interference pattern is obtained, even when the light intensity is so low that the interference pattern must be accumulated one photon at a time. The emergence of the interference pattern from individual photon events is illustrated in Fig. 3, in which we see the build-up of the two-slit interference pattern as single photon events (green points) are accumulated, one at a time. Based on Taylor's experimental results, in 1926 G. N. Lewis [18] reasoned, in a remarkable precursor to the Transactional Interpretation, that "an atom never emits light except to another atom …I propose to eliminate the idea of mere emission of light and substitute the idea of transmission, or a process of exchange of energy between two definite atoms or molecules."

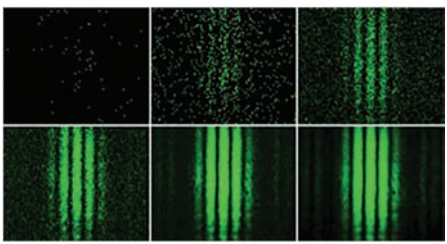

**Fig. 3: Build-up of a two-slit interference pattern in a Young's two-slit experiment at low illumination intensity as more and more single-photon events (green points) are accumulated [19].**

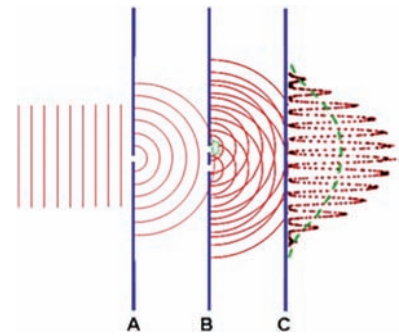

**Fig. 2: Young's two-slit experiment. Light waves diffract from the aperture in screen A, pass through two slits in screen B, and produce a "comb" interference pattern in their overlap region on screen C. The green/dashed line at C shows the diffraction pattern that would be observed if the two paths through the slits were made distinguishable, e.g., put in different states of polarization by a half-wave plate, shown behind the upper slit at B.**

The emergence of the interference pattern from individual photon events is the "quantum mystery" to which Richard Feynman referred: How is it possible that an ensemble of single photons, arriving at the screen one at a time, can produce such a wave-like interference pattern? It would appear that each individual photon particle must pass through both slits and must interfere with itself at the screen..

The Transactional Interpretation explains the puzzling build-up of a wave interference pattern from photon events as follows: in Fig. 2 the source emits plane offer waves moving to the right that are diffracted at screen **A**, pass through both slits at screen **B**, and arrive at any point on screen **C** from two directions. At locations along screen **C** where the two components of the offer wave interfere constructively there is a high probability of transaction formation, and at locations where the two components of the offer wave interfere destructively and cancel there is zero probability of a transaction.

Confirmation waves propagate to the left, moving back through the slits at **B** and the aperture at **A** to the light source. There the source, which is seeking to emit one photon, selects among the confirmation offers, and a transaction delivers a photon to screen **C**. The position at which the photon arrives is likely to be where the offer waves were constructive and unlikely to be where the waves were destructive. Therefore, an interference pattern that is made of many single photon transactions that build up on screen **C**, as shown in Fig. 3, is a natural consequence of the Transactional Interpretation.

The interference suppression from labeling can also be explained by the TI. Screen **C** receives offer waves that have passed through both slits and returns corresponding confirmation waves to the source. However, the vertically polarized offer wave will cause the return of a vertically polarized confirmation, and likewise for the horizontally polarized offer wave. The confirmation wave echo arriving at the source will only match the vertical polarization of the source if it returned through the same slit that the corresponding offer had passed through, so the transaction that forms will pass through only one of the two slits. Therefore, there will be no two-slit interference pattern for this case.

### 3.3 Wheeler's Delayed Choice Experiment (1978)

In 1978, John A. Wheeler raised another interpretational issue

[20] that is now known as Wheeler's Delayed-Choice Experiment (Fig. 4). Suppose that we have a Young's two slit interference apparatus as discussed above, with photons produced by a light source that illuminates two slits. The source emits one and only one photon in the general direction of the slits during the time interval chosen by the observer who is operating the apparatus. Downstream of the slits are two different measuring devices. One of these is a photographic emulsion **σ1** that, when placed in the path of the photons, will record photons' positions as they strike the emulsion, so that after many photon events, the emulsion will show a collection of spots that form a two-slit interference pattern characteristic of the photons' wavelength, momentum, and the slit separation. The other measuring device consists of a lens focusing the slit-images on photographic emulsion **σ2** at image points **1'** and **2'**. A photon striking either image point tells us that the photon had passed through the slit that is imaged at that position. Therefore, detection at **σ2** constitutes a determination of the slit (**1** or **2**) through which the photon passed.

Such an apparatus is often used to illustrate the wave-particle duality of light. The light waves that form the interference pattern on the emulsion must have passed through both slits of the apparatus in order to interfere at the emulsion, while the photon particles that strike the photographic emulsion **σ2** can have passed through only one slit – the one imaged by the lens **L** at image point **1'** or **2'**. The photographic emulsion **σ1** measures momentum (and wavelength) and the photographic emulsion **σ2** measures position, i.e., conjugate variables are measured. Thus, the two experimental measurements are "complimentary" in Bohr's sense. The Uncertainty Principle is not violated, however, because only one of the two experiments can be performed with a given photon. But Wheeler is not done yet.

The emulsion **σ1** is mounted on a fast acting pivot mechanism, so that on command it can almost instantaneously either be raised into position to intercept the photon from the source or rapidly dropped out of the way so that the photon can proceed to **σ2**. Thus when the emulsion **σ1** is up, we make an interference measurement requiring the photon to pass through both slits, and when the emulsion **σ1** is down, we make a position measurement requiring that the photon pass through only one slit.

Wheeler's innovative modification of this old *gedankenexperiment* is this: We wait until a time at which the photon has safely passed the slits but has not yet reached the emulsion apparatus **σ1**. Only at that time do we decide whether to place the **σ1** emulsion up or down. The decision is made after the photon must have passed through the slit system. Therefore, the photon has already emerged from the slit system when the experimenter decides whether it should be caused to pass through one slit (emulsion down) or both slits (emulsion up). Wheeler concluded that the delayed-choice experiment illustrated his paradigm about quantum mechanics: "*No phenomenon is a real phenomenon until it is an observed phenomenon*."

It might be argued that there would not really be time enough for a conscious observer to make the measurement decision. However, Wheeler has pointed out that the light source might be a quasar, and the "slit system" might be a foreground galaxy that bends the light waves around both sides by gravitational lensing. Thus, there would be a time interval of millions of years for the decision to be made, during which time the light waves from the quasar were in transit from the foreground galaxy to the observer. The delayed choice experiment, since it seems to determine the path of the photon after it has passed through the slit system, has been used as an illustration of retrocausal effects in quantum processes.

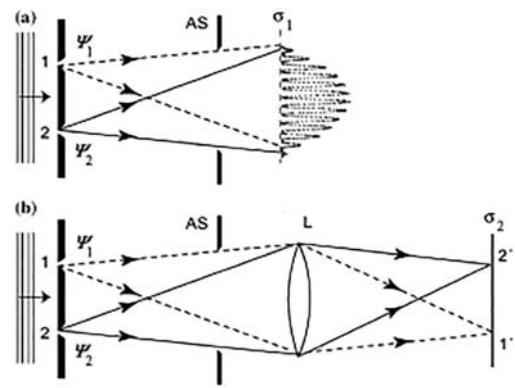

**Fig. 4: Wheeler's delayed choice experiment: Light from a single-photon source can either (a) produce an interference pattern on photographic emulsion σ1 or (b) be imaged by lens L to produce images of the two slits on photographic emulsion σ2 at points 1' and 2'. The experimenter waits until after the photon has passed through the slits to decide whether to lower photographic emulsion σ1 so that photographic emulsion σ2 provides which-slit information, or to leave it place so that the two-slit interference pattern characteristic of passage through both slits is observed at σ1.**

The *gedankenexperiment* does not lead to any explicit contradictions, but it demonstrates some of the retrocausal implications of the standard quantum formalism. In particular, the cause (emulsion **σ1** down or up) of the change in the photon's path has come after the effect (passage through one or two slits). There have been several experimental implementations of this experiment, the most recent (2007) performed by the Aspect Group in France [21]. All have shown the expected results, i.e., the predictions of standard quantum mechanics.

The Transactional Interpretation is able to give an account of the delayed choice experiment without resort to observers as collapse triggers. In the TI description, the source emits a retarded OW that propagates through slits **1** and **2**, producing offer waves $\psi_1$ and $\psi_2$. These reach the region of screen **σ1**, where either (a) they find the screen **σ1** up and form a two-path transaction with it as illustrated in Fig. 4(a) or; (b) they find the screen **σ1** down and proceed through lens **L** on separate paths to screen **σ2** where they strike the screen at image points **1'** and **2'** and create confirmation waves that return through the lens and slits to the source. In case (b), the source receives confirmation wave echoes from two separate sites on screen **σ2** and must decide which of them to use in a one-slit competed transaction, as shown by the solid and dashed lines in Fig. 4(b).

For case (a) in which the photon is absorbed by **σ1**, the advanced confirmation wave retraces the path of the OW, traveling in the negative time direction back through both slits and back to the source. Therefore the final transaction, as shown in Fig. 4(a), forms along the paths that pass through both slits in connecting the source with the screen **σ1**. The transaction is therefore a "two-slit" quantum event. The photon can be said to have passed through both slits to reach the emulsion.

For case (b) the offer wave also passes through both slits on its way to **σ2**. However, when the absorption takes place at one of the images (not both, because of the single quantum bound-ary condition), the lens focuses the confirmation wave so that it passes through only the slit imaged at the detection point. Thus the confirmation wave passes through only one slit in passing back from image to source, and the transaction which forms is characteristic of a "one-slit" quantum event. The source, receiving confirmation waves from two mutually exclusive one-slit possibilities, must choose only one of these for the formation of a transaction. The photon can be said to have passed through only one slit to reach σ2.

Since in the TI description the transaction forms atemporally, the issue of when the observer decides which experiment to perform is not significant. The observer determined the experimental configuration and boundary conditions and the transaction formed accordingly. Further, the fact that the detection event involves a measurement (as opposed to any other interaction) is not significant and so the observer has no special role in the process. To paraphrase Wheeler's paradigm, we might say: "No offer wave is a real transaction until it is a confirmed transaction".

### 3.4  The Afshar Experiment (2002)

The Afshar experiment [22] shows that, contrary to some of Niels Bohr's pronouncements about complementarity and wave particle duality, it is possible to see the effects of wave-like behavior and interference, even when particle-like behavior is being directly observed. In Bohr's words [23]: "…*we are presented with a choice of either tracing the path of the particle, or observing interference effects,… we have to do with a typical example of how the complementary phenomena appear under mutually exclusive experimental arrangements.*" In the context of a two-slit experiment, Bohr asserted [24] that complementarity in the Copenhagen Interpretation dictates that "*the observation of an interference pattern and the acquisition of which-way information are mutually exclusive.*"

The Afshar experiment, shown in Fig. 5, was first performed in 2003 by Shariar S. Afshar and was later repeated while he was a Visiting Scientist at Harvard. It used two pinholes in an opaque sheet illuminated by a laser. The light passing through the pinholes formed an interference pattern, a zebra-stripe set of maxima and zeroes of light intensity that were recorded by a digital camera. The precise locations of the interference minimum positions, the places where the light intensity went to zero, were carefully measured and recorded.

Behind the plane where the interference pattern formed, Afshar placed a lens that formed an image of each pinhole at a second plane. A light flash observed at image **1'** on this plane indicated unambiguously that a photon of light had passed through pinhole **1**, and a flash at image **2'** similarly indicated that the photon had passed through pinhole **2**. Observation of the photon flashes therefore provided particle path which-way information, as described by Bohr. According to the Copenhagen Interpretation, in this situation all wave-mode interference effects must be excluded. However, at this point Afshar introduced a new element to the experiment. He placed one or more vertical wires at the previously measured positions of the interference minima. In such a setup, if the wire plane was uniformly illuminated the wires absorbed about 6% of the light. Then Afshar measured the difference in the light intensity received at the pinhole image detectors with and without the wires in place.

We are led by the Copenhagen Interpretation to expect that when which-way information is obtained the positions of the interference minima should have no particular significance, and that the wires should intercept 6% of the light, as they do for uniform illumination. However, what Afshar observed was that the amount of light intercepted by the wires is very small, consistent with 0% interception. This implies that the interference minima are still locations of zero intensity and that the wave interference pattern is still present, even when which-way measurements are being made. Wires that are placed at the zero-intensity locations of the interference minima intercept no light. This observation would seem to create problems for the complementarity assertions of the Copenhagen Interpretation. Thus, the Afshar experiment is a significant quantum paradox.

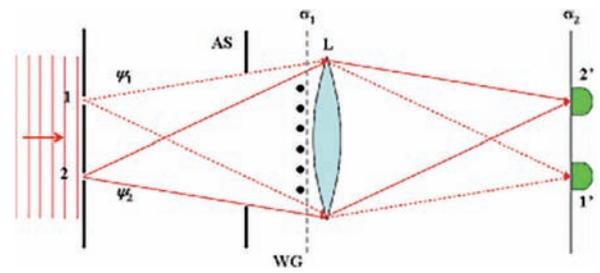

**Fig. 5:** In the Afshar experiment, a version of Wheeler's delayed-choice experiment is modified by placing vertical wires (WG) at the locations at which the interference pattern has interference minima on screen σ1. High transmission of light through the system when the wires are present and σ1 is absent implies that the interference pattern is *still* present, even when which-way information is available from the downstream detectors 1' and 2'.

The Transactional Interpretation explains Afshar's results as follows: The initial offer waves pass through both slits on their way to possible absorbers. At the wires, the offer waves cancel in first order, so that no transactions to wires can form, and no photons can be intercepted by the wires. Therefore, the absorption by the wires should be very small (<<6%) and consistent with what is observed. This is also what is predicted by the QM formalism. The implication is that the Afshar experiment has revealed a situation in which the Copenhagen Interpretation has failed to properly map the standard formalism of quantum mechanics.

We note that the many-worlds interpretation of quantum mechanics [25, 26] asserts that interference between its "worlds" (e.g., paths taken by particles) should not occur when the worlds are quantum-distinguishable. Therefore, the "many-worlds interpretation would also predict that there should be no interference effects in the Afshar experiment. Thus, the "many-worlds" interpretation has also failed to properly map the standard formalism of quantum mechanics.

### 3.5  The Freedman-Clauser EPR Experiment (1972)

Another quantum puzzle is the Freedman-Clauser experiment [27]. An atomic 2-photon cascade source produces a pair of polarization-entangled photons. If we select only entangled photons emitted back-to-back, then because of angular momentum conservation, both photons must be in the same state of circular or linear polarization. Measurements on the photons with linear polarimeters in each arm of the experiment show that when the planes of the polarimeters are aligned, independent of the direction of alignment, the two polarimeters always measure HH or VV for the two linear polarization states, i.e.,

J.C. Cramer

both photons are always in the same linear polarization state.

When the polarization plane of one polarimeter is rotated by an angle θ with respect to the other polarization plane, some opposite-correlation **HV** and **VH** events creep in. If θ is increased, the fraction of these events grows proportional to **1−cos2(θ)**, which for small values of θ is proportional to **θ2**. This polarization correlation behavior produces a dramatic violation of the Bell inequalities [28], which for local hidden variable alternatives to standard quantum mechanics require a growth in **HV** and **VH** events that is linear with θ. The implication of the Bell-inequality violations is that quantum nonlocality is required to explain the observed quadratic polarization correlations.

How are the nonlocality-based polarization correlations of the Freedman–Clauser experiment possible? The Transactional Interpretation provides a clear answer, which is illustrated in Fig. 6. The source of the polarization-entangled photons seeks to emit the photon pair by sending out offer waves $\psi_L$ and $\psi_R$ to the left and right detectors.

The detectors respond by returning confirmation waves $\psi_L*$ and $\psi_R*$ back to the source. A completed three-vertex transaction can form from these echoes, however, only if the two potential detections are compatible with the conservation of angular momentum at the source. This requirement produces the observed polarization correlations. The transaction does not depend on the separation distance of the polarimeters or on which of the polarization detection events occurs first, since the transaction formation is atemporal, and it even-handedly treats any sequence of detection events.

### 3.6   Interaction-Free Measurements (1993)

In 1993, Elitzur and Vaidmann [29] (EV) showed a surprised physics community that quantum mechanics permits the non-classical use of light to examine an object without a single photon of the light actually interacting with the object. The EV experiment requires only the possibility of an interaction.

In their paper [29] Elitzur and Vaidmann discuss their scenario in terms of the standard Copenhagen Interpretation of quantum mechanics, in which the interaction-free result is rather mysterious, particularly since the measurement produces "knowledge" that is not available classically. They also considered their scenario in terms of the Everett–Wheeler or "many-worlds" interpretation of quantum mechanics [25, 26].

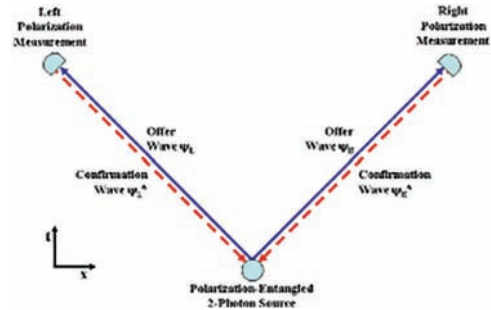

**Fig. 6:** Spacetime schematic of a nonlocal "V" transaction for visualizing the polarization-entangled Freedman–Clauser EPR experiment. Offer waves $\psi_L$ and $\psi_R$ (blue/solid) move from source to linear polarization detectors, and in response, confirmation waves $\psi_L*$ and $\psi_R*$ (red/dashed) move from detectors to source. The three-vertex transaction can form only if angular momentum is conserved by having correlated and consistent measured linear polarizations for both detected photons.

Considering the latter, they suggest that the information indicating the presence of the opaque object can be considered to have come from an interaction that had occurred in a separate Everett–Wheeler universe and was transferred to our universe through the absence of interference. Here we will examine the same scenario in terms of the Transactional Interpretation and will provide a more plausible account of the physical processes that underlie interaction-free measurements.

The basic apparatus used by EV is a Mach–Zender interferometer, as shown in Fig. 7. Light from a light source L goes to a 50:50% beam splitter $S_1$ that divides incoming light into two possible paths or beams. These beams are deflected by 90° by mirrors A and B, so that they meet at a second beam splitter $S_2$, which recombines them by another reflection or transmission. The combined beams from $S_2$ then go to the photon detectors $D_1$ and $D_2$.

The Mach–Zehnder interferometer has the characteristic that, if the paths A and B have precisely the same path lengths, the superimposed waves from the two paths are in phase at $D_1$ ($\Delta\varphi = 0$) and out of phase at $D_2$ ($\Delta\varphi = \pi$). This is because with beam splitters, an emerging wave reflected at 90° is always 90° out of phase with the incident and transmitted waves [30]. The result is that all photons from light source L will go to detector $D_1$ and none will go to detector $D_2$.

Now, as shown in Fig. 8, we place an opaque object (Obj) on path A. It will block light waves along the lower path after reflection from mirror A, ensuring that all of the light arriving at beam splitter $S_2$ has traveled there via path B. In this case there is no interference, and beam splitter $S_2$ sends equal components of the incident wave to the two detectors.

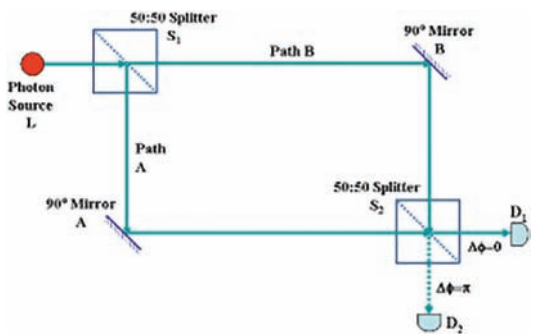

**Fig. 7:** Mach Zehnder interferometer with both beam paths open. All photons go to $D_1$ because of destructive interference at $D_2$.

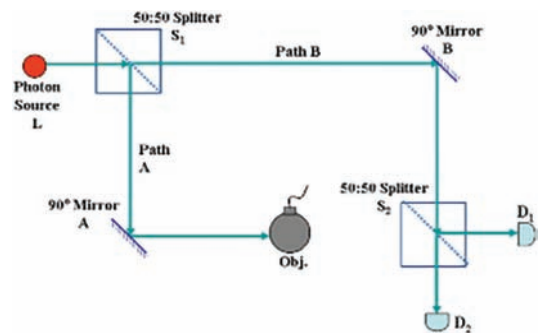

**Fig. 8:** Mach Zehnder interferometer with one beam path blocked. Half of the photons are absorbed by the blocking object, 25% go to $D_1$, and 25% go to $D_2$.

Now suppose that we arrange for the light source L to emit only one photon within a given time period. Then, if we do the measurement with no opaque object on path A, we should detect the photon at $D_1$ 100% of the time. If we perform the same measurement with the opaque object Obj blocking path A, we should detect a photon at $D_1$ 25% of the time, a photon at $D_2$ 25% of the time, and should detect no photon at all 50% of the time (because it was removed by Obj in path A). In other words, the detection of a photon at $D_2$ guarantees that an opaque object is blocking path A, although no photon had actually interacted with object Obj. This is the essence of the Elitzur and Vaidmann interaction-free measurement.

Note that if a photon is detected at detector $D_1$, the issue of whether an object blocks path A is unresolved. However, in that case another photon can be sent into the system, and this can be repeated until either a photon is detected at $D_2$ or absorbed by Obj. The net result of such a recursive procedure is that 66% of the time a photon will strike the object, resulting in no detection signal, while 33% of the time a photon will be detected at $D_2$, indicating without interaction that an object blocks the A path. Thus, the EV procedure has an efficiency of 33% for non-interactive detection.

As before, in analyzing interaction-free measurements with the Transactional Interpretation, we will explicitly indicate offer waves ψ by a specification of the path in a Dirac ket state vector ψ =| path>, and we will underline the symbols for optical elements at which a reflection has occurred. Confirmation waves ψ∗ will similarly be indicated by a Dirac Bra state vector ψ∗ = <path |, and will indicate the path considered by listing the elements in the time-reversed path with reflections underlined.

Consider first the situation in which no object is present in path A as shown in Fig. 8. The offer waves from L to detector $D_1$ are | L-$S_1$-A-$S_2$-$D_1$> and | L-$S_1$-B-$S_2$- $D_1$>. They arrive at detector $D_1$ in phase because the offer waves on both paths have been transmitted once and reflected twice. The offer wave from L initially has unit amplitude, but the splits at $1/\sqrt{2}$ each reduce the wave amplitude by $1/\sqrt{2}$ so that each wave, having been split twice, has an amplitude of ½ as it reaches detector $D_1$. Therefore, the two offer waves of equal amplitude and phase interfere constructively, reinforce, and produce a confirmation wave that is initially of unit amplitude.

Similarly, the offer waves from L to detector $D_2$ are | L-$S_1$-A-$S_2$-$D_2$> and | L-$S_1$-B-$S_2$- $D_2$>. They arrive at detector D2 180° out of phase, because the offer wave on path A has been reflected three times while the offer wave on path B has been transmitted twice and reflected once. Therefore, the two waves with amplitudes ±i/2 interfere destructively, cancel at detector $D_2$, and produce no confirmation wave. The confirmation waves from detector $D_1$ to L are <$D_1$-$S_2$-A-$S_1$-L | and <$D_1$-$S_2$-B-$S_1$- L |. They arrive back at the source L in phase because, as in the previous case, the confirmation waves on both paths have been transmitted once and reflected twice.

As before the splits at $S_1$ and $S_2$ each reduce the wave amplitude by $1/\sqrt{2}$, so that each confirmation wave has an amplitude of ½ as it reaches source L. Therefore, the two offer waves interfere constructively, reinforce and have unit amplitude. Since the source L receives a unit amplitude confirmation wave from detector $D_1$ and no confirmation wave from detector $D_2$, the transaction forms along the path from L to $D_1$ via A and B. The result of the transaction is that a photon is always transferred from the source L to detector $D_1$ and that no photons can be transferred to $D_2$. Note that the transaction forms along both paths from L to $D_1$. This is a transactional account of the operation of the Mach–Zender interferometer.

Now let us consider the situation when the object blocks path A as shown in Fig. 9. The offer wave on path A is | L-$S_1$-A-Obj>. As before an offer wave on path B is | L-$S_1$-B-$S_2$-$D_1$>, and it travels from L to detector $D_1$. The wave on path B also splits at $S_2$ to form offer wave | L-$S_1$-B-$S_2$-$D_2$>, which arrives at detector $D_2$. The splits at $S_1$ and $S_2$ each reduce the wave amplitude by $1/\sqrt{2}$, so that the offer wave at each detector, having been split twice, has an amplitude of ½ . However, the offer wave | L-$S_1$-A-Obj> to the object in path A, having been split only once, is stronger and has amplitude of $1/\sqrt{2}$.

In this situation, the source L will receive confirmation waves from both detectors and also from the object. These, respectively, will be confirmation waves <$D_1$-$S_2$-B-$S_1$- L |, <$D_2$-$S_2$-B-$S_1$-L | and <Obj-A-$S_1$-L |. The first two confirmation waves started from their detectors with amplitudes of ½ (the final amplitude of their respective offer waves) and have subsequently been split twice. Therefore, they arrive at source L with amplitudes of ¼. On the other hand, the confirmation wave from the object initially has amplitude $1/\sqrt{2}$ , and it has been split only once, so it arrives at the source with amplitude ½ .

The source L has one photon to emit and three confirmations to choose from, with round-trip amplitudes (ψψ∗) of ¼, to $D_1$ ¼ to $D_2$, and ½ to object Obj. In keeping with the probability assumption of the Transactional Interpretation and Born's probability law, it will choose with a probability proportional to these amplitudes. Therefore, the emitted photon goes to $D_1$ 25% of the time, to $D_2$ 25% of the time, and to object Obj in path A 50% of the time. As we have seen above, the presence of the object in path A modifies the detection probabilities so that detector $D_2$ will receive ¼ of the emitted photons, rather than none of them, as it would do if the object were absent.

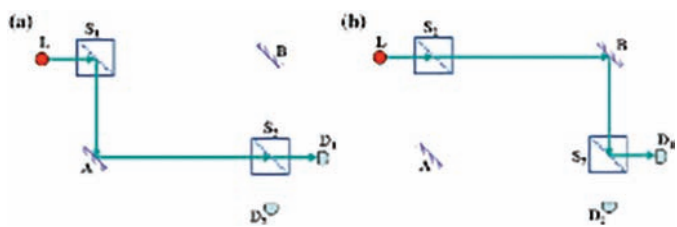

**Fig. 9:** Offer waves (a) | L-$S_1$-A-$S_2$-D1> and (b) | L-$S_1$-B-$S_2$-D1>.

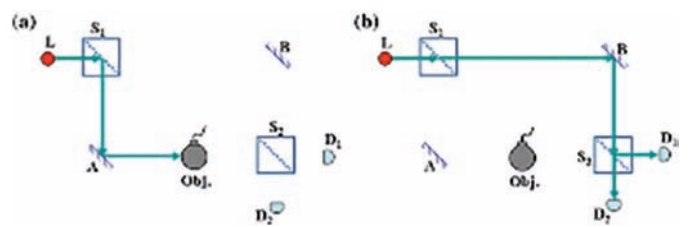

**Fig. 10:** Offer waves (a) | L-$S_1$-A-Obj> and (b) | L-$S_1$-B-$S_2$-D1> + | L-$S_1$-B-$S_2$-D2>.

How can the transfer of non-classical knowledge be understood in terms of the transactional account of the process? In the case where there is an object in the A path, it is probed both by the offer wave from L and by the aborted confirmation waves from $D_1$ and $D_2$. The latter are 180° out of phase and cancel. When we detect a photon at $D_2$, (i.e., when a transaction forms between L and $D_2$), the object has not interacted with a photon (i.e., a transaction has not formed between L and the object Obj). However, it has been probed by an offer wave from the source, which "feels" its presence and modifies the interference balance at the detectors, providing non-classical information. Thus, the Transactional Interpretation gives a simple explanation of the mystery of interaction-free measurements.

### 3.8 The Hardy One-Atom *Gedankenexperiment* (1992)

In 1992 Lucien Hardy [31, 32] proposed the *gedankenexperiment* shown in Fig. 11, which is a modified version of the interaction-free measurement scenario of Elitzur and Vaidmann [29] (see Sect. 6.13) in which their blocking object (or bomb) is replaced by a single spin-½ atom, initially prepared in an X-axis +½ spin-projection, then Stern–Gerlach separated [33] into one of two spatially separated boxes that momentarily contain the atom in its Z-axis +½ and −½ spin projections, then transmit their contents to be recombined by an inverse Stern–Gerlach process, so that the X-axis projection of the atom can be measured.

The Z-spin +½ box (Z+) is placed directly in one path of a Mach–Zehnder interferometer, so that if the atom is present in that box during photon transit, it has a 100% probability of absorbing a photon traveling along that arm of the interferometer. After a single photon from light source L traverses the interferometer, the final X-axis spin projection of the atom is measured. The non-classical outcome of the *gedankenexperiment* is that, for events in which a photon is detected by dark detector D, the spin measurement of the atom has a 50% probability of having an X-axis spin projection of −½, even though the atom had previously been prepared in the +½ X-axis spin state, and the atom had never directly interacted with the photon.

Hardy analyzes the measurement in terms of the Bohm–de Broglie interpretation/revision of quantum mechanics [34] and concludes that the non-classical outcome of the measurement can be attributed to "empty waves", by which he means de Broglie guide waves that have traversed the interferometer along paths not subsequently followed by the single emitted photon. At least four other papers [35-38] have analyzed the Hardy *gedankenexperiment* using alternative QM interpretations that focus on wave function collapse, notably the "collapse" and the "consistent histories" interpretations.

The Transactional Interpretation explains the transfer of non-classical knowledge in terms of the transactional account of the process. In particular, in the case where there is an atom in the **v** path, it is probed by the offer wave from **L**. When we detect a photon at **D**, (i.e., when a transaction forms between **L** and **D**), the object has not interacted with a photon (i.e., a transaction has not formed between **L** and the atom in box Z+). However, the atom has been probed by offer waves from **L**, which "feel" its presence and modify the interference balance at the detectors and the spin statistics of the atom. Thus, the Transactional Interpretation gives a simple explanation of the Hardy *gedankenexperiment*.

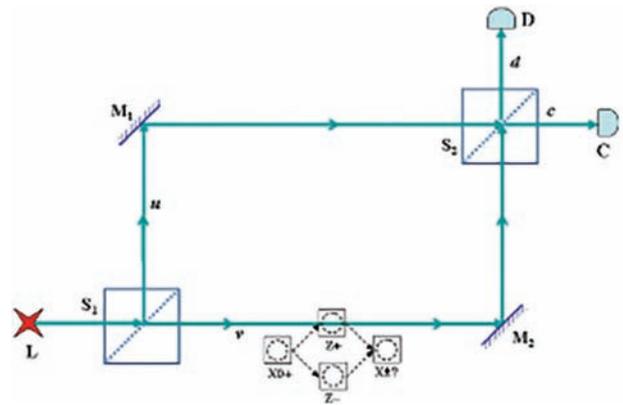

**Fig. 11:** The Hardy single-atom interaction-free measurement.

### 3.9 The Quantum Eraser (1995)

A more elaborate delayed-choice variation is the quantum eraser experiment, a high-tech descendant of Wheeler's delayed choice concept. The experiment used a new (in 1995) trick for making "entangled" quantum states. If ultraviolet light from a 351 nanometer (nm) argon-ion laser passes through a LiIO3 crystal, non-linear effects in the crystal can "split" the laser photon into two longer wavelength photons at 633 nm and 789 nm in a process called "down-conversion". The energies of these two "daughter" photons add up to the energy of their pump-photon parent, as do their vector momenta, and they are connected non-locally because they constitute a single "entangled" quantum state. They are required to be in correlated states of polarization, and under the conditions of this down-conversion they will be vertically polarized. As in other EPR experiments, a measurement performed on one of these photons affects the outcome of measurements performed on the other.

In a version of the experiment performed by Anton Zeilinger's group in Innsbruck, Austria, [39] the laser beam is reflected so that it makes two passes through the nonlinear crystal, so that an entangled photon pair may be produced in either the first or the second pass through the non-linear crystal. As shown in Fig. 12, the experiment has the configuration of a six-pointed star formed of three beam paths intersecting at a point inside the crystal. The laser beam first passes through the crystal moving horizontally downstream, is reflected by a downstream mirror $\Phi_P$, and then passes through the crystal again moving horizontally upstream. Along the two diagonal branches downstream of the laser the two down-converted photons made in the first laser-pass travel to mirrors $\Phi_S$ and $\Phi_I$ (S for signal and I for idler), where they are reflected back to their production point and travel past it to upstream detectors $D_S$ and $D_I$. The laser beam, in making its second pass through the crystal has a second chance to make a pair of down-converted photons. If these are produced, they travel directly to the upstream detectors along the two upstream diagonal branches.

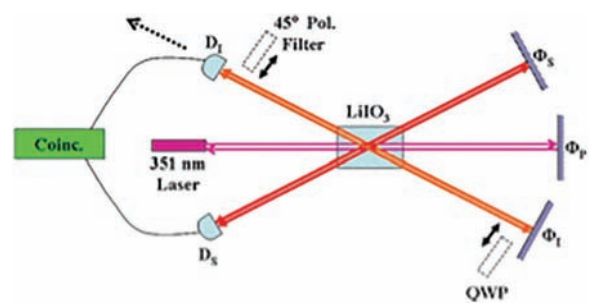

**Fig. 12:** Schematic diagram of the quantum eraser experiment. A LiIO3 nonlinear crystal is pumped by a 351nm laser beam (violet) and produces by down-conversion vertically polarized 633 nm (orange) and 789 nm (red) photons that can be made in either pump-photon pass through the crystal. A quarter-wave plate (QWP) and 45° polarizing filter may be inserted in the I path and the path to $D_I$ may be lengthened to produce a time delay (see text).

The net result is that a photon arriving in coincidence at the two upstream detectors may have been produced in either the first laser pass through the crystal and then reflected to the detector, or in the second pass and traveled directly to the detector. There is no way of determining which "history" (direct vs. reflected) happened, so the states are superimposed. Therefore, the quantum wave functions describing these two possible production histories must interfere. The interference may be constructive or destructive, depending on the interference phase determined by the downstream path lengths (all about 13 cm) to the three mirrors of the system. Changing the path length to one of the mirrors (for example, by moving the laser-beam reflector $\Phi_P$) is observed to produce a succession of interference maxima and minima in the two detectors.

This experimental setup is governed by the same physics as the delayed-choice experiment of Sect. 3.3, but, because there are two coincident photons and well separated paths for the two possible histories, it is easier to play quantum tricks with the system. Initially, all polarizations are vertical. Now the experiment is modified to remove the quantum interference by placing distinguishing polarization labels on the two possible photon histories (direct vs. reflected). A transparent optical element called a "quarter-wave plate" (QWP) is placed in front of the photon reflection mirror $\Phi_I$. The QWP is set to rotate the polarization state of the reflected photons from vertical to horizontal polarization as they pass twice through it. This polarization modification allows the reflected and direct "histories" to be quantum-distinguishable, because one of the reflected photons is horizontally polarized while the direct photons are vertically polarized. The two superimposed quantum states are now distinguishable (even if no polarization measurement is actually made), and the interference pattern is eliminated, both in the I arm of the experiment in which the QWP is placed and also in the other S arm, where no modification was made.

Finally, the "quantum eraser" is brought into use. Any vertically or horizontally polarized light beam can be separated into a light component polarized 45° to the left of vertical and a light component polarized 45° to the right of vertical. Therefore, for the photons with the QWP in front of their mirror, placing just in front of their detector a filter that passes only light polarized 45° to the left of vertical "erases" the label that had distinguished the two histories by making the polarizations of the two waves reaching detector $D_I$ the same. When this is done, it is found that interference is restored.

Further, the paths to the two detectors can have different lengths, with the path through the 45° filter to $D_I$ made much longer than the path to detector $D_S$. This has the effect of erasing the path-distinguishing label on the I photon after the S photon had already been detected. This modification is observed to have no effect on the interference. The *post-facto* erasure still restores interference. The path label can be erased retroactively and has the same effect (retroactive or not) on the quantum interference of the waves. Effectively, the quantum eraser has erased the past!

The Transactional Interpretation can easily explain the curious retroactive erasure of "which-way" information. When which-way information is present, separate transactions must form for each of the paths, and no interference can be observed. When the which-way information is erased, the overall transaction that forms involves both paths, and interference is observed. Modifying the polarizations causes a different type of transaction formation, resulting in different observations. The retroactive erasure of the which-way information is irrelevant, because the transaction forms atemporally, connecting the source and detectors in single or double advanced-retarded TI handshakes across spacetime.

### 3.10 The Black Hole Information Paradox (1975-2015)

Stephen Hawking's 1975 calculations [40] predicting black hole evaporation by Hawking radiation described a process that apparently does not preserve information. This created the Black Hole Information Paradox, which has been an outstanding problem at the boundary between general relativity and quantum mechanics ever since. Lately, gravitational theorists have focused on pairs of quantum-entangled particles, in part because the particle pair involved in Hawking radiation should be entangled. They have considered ways in which the quantum entanglement might be broken or preserved when one photon of the entangled photon pair crosses the event horizon and enters a black hole.

One recent suggestion is that the quantum entanglement breaks (whatever that means) when the infalling member of the entangled particle pair crosses the event horizon, with each breaking link creating a little burst of gravitational energy that cumulatively create a firewall just inside the event horizon. This firewall then destroys any infalling object in transit [41]. The firewall hypothesis, however, remains very controversial, and there is no apparent way of testing it.

More recently Maldacena and Susskind [42] have suggested an alternative. When two entangled black holes separate, they hypothesize that a wormhole connection forms between them to implement their entanglement. It has even been suggested that such quantum wormholes may link all entangled particle pairs. There are, however, problems with this interesting scenario, not the least of which is that such wormholes should have significant mass that is not observed.

The Transactional Interpretation offers a milder, if less dramatic solution to this problem, providing an interesting insight into the Black Hole Information Paradox. One normally thinks that absolutely nothing can break out of the event horizon of a black hole from the inside and escape. However, as illustrated in Fig. 13, there is one exception: advanced waves can emerge from a black hole interior, because they are just the time-reverse of a particle-wave falling in. An advanced wave "sees" the black hole in the reverse time direction, in which it looks like a white hole that emits particles. The strong gravitational force facilitates rather than preventing the escape of an advanced wave. Thus, an entangled particle pair, linked by an advanced-retarded wave handshake, has no problem in maintaining the entanglement, participating in transactions, and preserving conservation laws, even when one member of the pair has fallen into a black hole. There is no need for entanglement-breaking firewalls or entanglement-preserving wormholes, just a trans-

J.C. Cramer

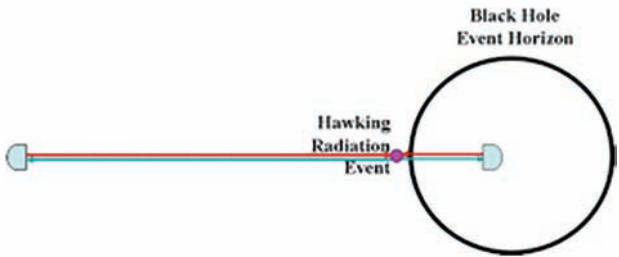

**Fig. 13:** Transaction through Black Hole event horizon using escaping advanced waves.

actional handshake. Thus, it would seem that the Transactional Interpretation goes some considerable distance toward solving the Black Hole Information Paradox and resolving an issue that divides quantum mechanics and gravitation and providing a mechanism for preserving information across event horizons.

## 4 THE PROCESS OF FORMING TRANSACTIONS

Some critics of the Transactional Interpretation have asked why it does not provide a detailed mathematical description of transaction formation. This question betrays a fundamental misunderstanding of what an interpretation of quantum mechanics actually is. In our view, the mathematics is (and should be) exclusively contained in the standard quantum formalism itself. The function of the interpretation is to interpret that mathematics, not to introduce any new additional mathematics. We note, however, that this principle is violated by the Bohm-de Broglie "interpretation", in the Ghirardi-Rimini-Weber "interpretation", and in many other so-called interpretations. In that sense, these are not interpretations of standard quantum mechanics at all, but rather are alternative theories.

It is true that while the Transactional Interpretation leans heavily on the quantum formalism, the standard formalism of quantum mechanics does not contain mathematics that explicitly describes wave function collapse (which the TI interprets as transaction formation). However, there has been an application of the standard QM formalism in the literature that provides a detailed mathematical description of the "quantum-jump" exponential build-up of a transaction involving the transfer of a photon from one atom to another. In particular, Carver Mead does this in Sect. 5.4 of his book *Collective Electrodynamics* [43].

Briefly, Mead considers an emitter atom in an excited state with excitation energy $E_1$ and a space-antisymmetric wave function of $\psi_E = A_E(r) \exp(-i E_1 t/\hbar)$ and a structurally identical absorber atom in its ground state with excitation energy $E_0$ and a space-symmetric wave function of $\psi_A = S_A(r) \exp(-i E_0 t/\hbar)$, where $A$ is an antisymmetric function and $S$ is a symmetric function. Both of these are stable states with no initial dipole moments. He assumes that the initial positive-energy offer wave from the excited emitter atom **E** interacting with the absorber atom **A** perturbs it into a mixed state that adds a very small component of excited-state wave function to its ground-state wave function. Similarly, the negative-energy confirmation wave echo from the absorber atom interacting with the emitter atom perturbs it into a mixed state that adds a very small component of ground-state wave function to its excited-state wave function, as shown schematically in Fig. 14.

Because of these perturbations, both atoms develop small time-dependent dipole moments that, because of the mixed-energy states, oscillate with the same beat frequency $\omega = (E_1 - E_0)/\hbar$ and act as coupled dipole resonators. The phasing of their resulting waves is such that energy is transferred from emitter to absorber at a rate that initially rises exponentially.

To quote from Mead's discussion:

*The energy transferred from one atom to another causes an increase in the minority state of the superposition, thus increasing the dipole moment of both states and increasing the coupling and, hence, the rate of energy transfer. This self-reinforcing behavior gives the transition its initial exponential character.*

In other words, Mead shows mathematically that the perturbations induced by the initial offer/confirmation exchange trigger the formation of a full-blown transaction in which a photon-worth of energy $E_1 - E_0$ is transferred from emitter to absorber, resulting in a confirmation wave from absorber similarly perturbing the emitter. The result is a pair of dipole resonators oscillating at the same beat frequency $\omega = (E_1 - E_0)/\hbar$, to produce an exponentially rising coupling and transaction formation. Thus, mutual offer/confirmation perturbations of the emitter and absorber acting on each other create a frequency-matched pair of dipole resonators as mixed states, and this dynamically unstable system must either exponentially avalanche to the formation of a completed transaction or disappear when a competing transaction forms.

In a universe full of particles, this process does not occur in isolation, and both emitter and absorber are also randomly perturbed by waves from other systems that can randomly drive the exponential instability in either direction. This is the source of the intrinsic randomness in quantum processes – the missing random element that changes quantum mechanics from the determinism of classical mechanics. Ruth Kastner [44] likes to describe this intrinsic randomness as "spontaneous symmetry breaking", which perhaps clarifies the process by analogy with quantum field theory.

Because the waves carrying positive energy from emitter to absorber are retarded waves with positive transit time, and the waves carrying negative energy from absorber to emitter are advanced waves with negative transit time, there is no net time delay – aside from time-of-flight propagation time of the transferred energy, in the quantum-jump process – and it is effectively instantaneous. Thus, the Transactional Interpretation explains Niels Bohr's "instantaneous" quantum jumps – a concept that Schrödinger found impossible to accept [45].

In Figs. 15 to 19 we have used Mead's formalism with standard hydrogen-atom wave functions to calculate example transactions and to make plots of various aspects of the transaction formation in progress.

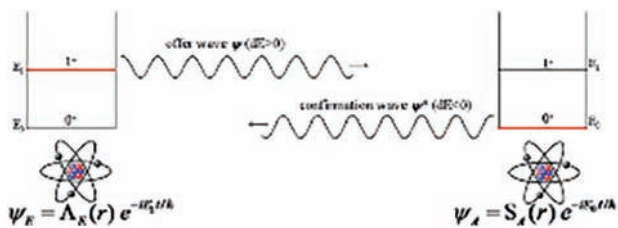

**Fig. 14:** Mead model of transaction formation: Emitter in antisymmertric excited state of energy E1 perturbs absorber in symmetric ground state of energy E0 with offer wave.

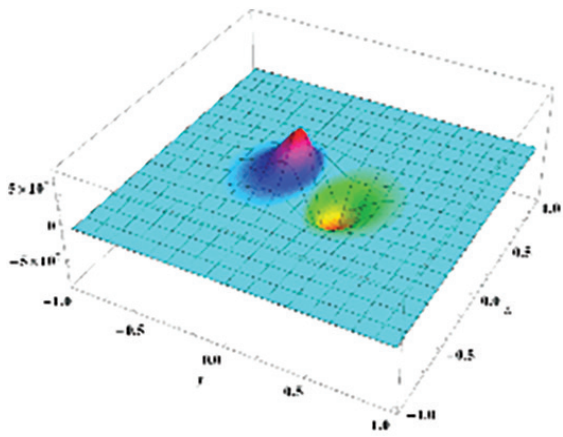

Fig. 15: Electric dipole oscillating at beat frequency $\omega=(E_1-E_0)/\hbar$ created by mixture of states with excitation energies $E_0$ and $E_1$.

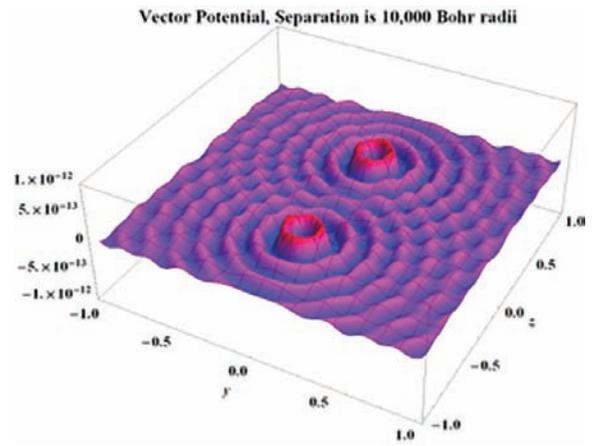

Fig. 16: Waves of electric potential created by coupled dipole oscillations in atoms undergoing energy exchange transaction.

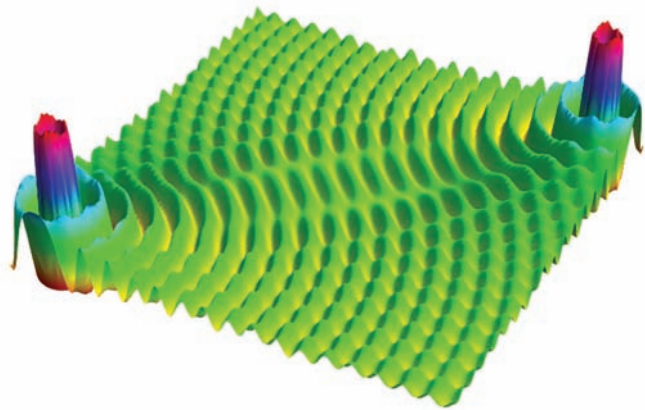

Fig. 17: Waves of electric potential created by coupled dipole oscillations in atoms undergoing energy exchange transaction.

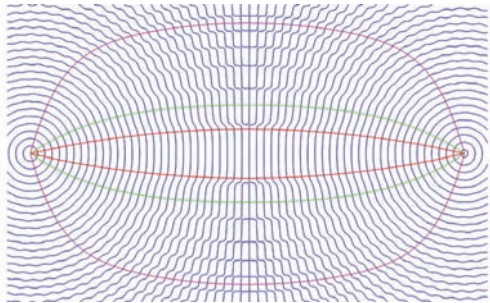

Fig. 18: Paths of equal phase from emitter atom, arriving at absorber atom to coherently reinforce the developing transaction.

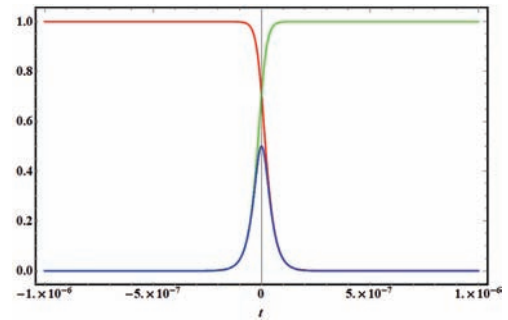

Fig. 19: Amplitudes of the excited state and ground state wave functions that are present in the emitter atom as it is undergoing a transaction.

This is, of course, not a general proof that the offer/confirmation exchange always triggers the formation of a transaction, but it represents a demonstration of that behavior in a tractable case, and it represents a prototype of the general transaction behavior. It further demonstrates that the transaction model is implicit in and consistent with the standard quantum formalism, and it demonstrates how the transaction, as a spacetime standing wave connecting emitter to absorber, can form.

## 5   CONCLUSIONS

- The Transactional Interpretation provides a rational way of visualizing and understanding the mechanisms behind entanglement, nonlocality, and wave function collapse.

- The plethora of interpretational paradoxes and non-classical quantum-optics experimental results can all be understood by applying the Transactional Interpretation.

- The process of transaction formation, at least in simple cases, emerges directly from the application of standard quantum mechanics to the advanced-retarded-wave handshake process as it builds and avalanches to completion.

- As the mattress commercial asks: ***Why buy your Quantum Interpretation anywhere else?***